\RequirePackage{xspace}
\documentclass{ws-procs9x6}
\newcommand\Acp {\ensuremath{A_{\CP}}}
\def\babar{\mbox{\sl B\hspace{-0.4em} {\small\sl A}\hspace{-0.37em} \sl B\hspace{-0.4em} {\small\sl A\hspace{-0.02em}R}}}

\def\ctwob{\ensuremath{\cos\! 2 \beta   }\xspace}
\def\btoccbars{\ensuremath{\b\rightarrow\c\cbar\s}}
\def\pep2{PEP-II}
\def\epem{\ensuremath{e^+e^-}\xspace}
\newcommand{\UfourS}{\ensuremath{\Upsilon(4S)}}
\def\CP{\ensuremath{C\!P}\xspace}
\def\B{\ensuremath{B}\xspace}
\def\stwob{\ensuremath{\sin\! 2 \beta   }\xspace}
\def\ra                 {\ensuremath{\rightarrow}\xspace}
\def\Bzb     {\ensuremath{\Bbar^0}\xspace}
\def\Bbar    {\kern 0.18em\overline{\kern -0.18em B}{}\xspace}
\def\Bz      {\ensuremath{B^0}\xspace}
\def\deltamd{\ensuremath{{\rm \Delta}m_d}\xspace}
\def\BB      {\ensuremath{B\Bbar}\xspace} 
\def\eps{\varepsilon\xspace}
\def\mistag{\ensuremath{w}\xspace}
\def\mes        {\mbox{$m_{\rm ES}$}\xspace}
\newcommand{\pvec}{{\bf p}}
\newcommand{\DE}{\ensuremath{\Delta E}}
\newcommand{\half}{\ensuremath{{1\over2}}}

\def\s     {\ensuremath{s}\xspace}

\def\c     {\ensuremath{c}\xspace}
\def\cbar  {\ensuremath{\overline c}\xspace}

\def\b     {\ensuremath{b}\xspace}

\def\jpsi     {\ensuremath{{J\mskip -3mu/\mskip -2mu\psi\mskip 2mu}}\xspace}
\def\KS    {\ensuremath{K^0_{\scriptscriptstyle S}}\xspace} 
\def\KL    {\ensuremath{K^0_{\scriptscriptstyle L}}\xspace} 
\newcommand{\etapr}{\ensuremath{\eta^{\prime}}\xspace}
\newcommand{\msp}{\ensuremath{\phantom{-}}}
\def\gsim{{~\raise.15em\hbox{$>$}\kern-.85em
          \lower.35em\hbox{$\sim$}~}\xspace}
\def\Dstarp  {\ensuremath{D^{*+}}\xspace}
\def\Dstarm  {\ensuremath{D^{*-}}\xspace}
\newcommand{\jprl}      [1]  {\jprlBase\ {\bf #1}}
\newcommand{\jprlBase}       {Phys.\ Rev.\ Lett.\xspace}
\newcommand{\nima}      [1]  {\nimBaseA~{\bf #1}}
\newcommand{\nimBaseA}       {Nucl.\ Instrum.\ Methods Phys.\ Res., Sect.\ A\xspace}
\newcommand{\jprd}      [1]  {\jprBase\ D~{\bf #1}}
\newcommand{\jprBase}        {Phys.\ Rev.\xspace}
\newcommand\etal{{\it et al.}}
\newcommand{\plb}       [1]  {\jplBase\ B~{\bf #1}}
\newcommand{\jplBase}        {Phys.\ Lett.\xspace}
\newcommand{\npBase}         {Nucl.\ Phys.\xspace}
\newcommand{\npb}       [1]  {\npBase\ B~{\bf #1}}

\begin{document}

\title{Measurements of CKM angle $\beta$ from \babar}

\author{K.~A. ULMER \\ representing the \babar\ Collaboration}

\address{Department of Physics \\ 
University of Colorado, \\
Boulder, CO 80309\\
E-mail: ulmerk@colorado.edu}  

\maketitle

\abstracts{
We present recent results of hadronic $B$ meson decays related to
the CKM angle $\beta$. The data used were collected by the \babar\
detector at the \pep2\ asymmetric-energy \epem\ collider operating
at the \UfourS\ resonance located at the Stanford Linear Accelerator
Center.
\\ \\ \babar-PROC-07/002 \\SLAC-PUB-12519
}

\section{Introduction}
The Standard Model (SM) of particle physics describes charge conjugation-parity 
(\CP) violation as a consequence of a complex phase in the three-generation 
Cabibbo-Kobayashi-Maskawa (CKM) quark-mixing matrix\cite{ref:ckm}. \CP
violation in $B$ meson decays is described by the angles $\alpha$, $\beta$ and
$\gamma$ of the Unitarity Triangle. We describe here recent results from \babar\ for the
angle $\beta$, defined as $\arg \left[\, -V_{\rm cd}^{}V_{\rm cb}^* / 
V_{\rm td}^{}V_{\rm tb}^*\, \right]$ where the $V_{ij}$ are CKM matrix elements.

The \babar\ detector\cite{ref:babar} is located at the SLAC PEP-II $e^+e^-$ 
asymmetric-energy \B -factory\cite{ref:pep}. Data
are collected at the $\UfourS$ resonance. 

These proceedings describe measurements of \stwob\ from $B^0\ra c\bar{c}K^0$ decays,
\stwob\ from loop-dominated charmless $B^0$ decays and measurements of \ctwob.

\section{Analysis Technique}
The angle $\beta$ is extracted through measurements of time-dependent $\CP$-asymmetries\cite{ref:babarprd}. $\Acp$ is defined as
\begin{eqnarray}\label{eq:acp}
\Acp(t) & \equiv & \frac{N(\Bzb(t)\to f) - N(\Bz(t)\to f)} {N(\Bzb(t)\to f) + N(\Bz(t)\to f)} \nonumber \\
        &&{} = S \sin(\deltamd{t}) - C \cos(\deltamd{t}),
\label{eq:timedependence}
\end{eqnarray}
where $N(\Bzb(t)\to f)$ is the number of \Bzb\ that decay into the $CP$-eigenstate $f$ after a time $t$
and $\deltamd$ is the difference between the \B\ mass eigenstates.
The sinusoidal term describes interference between $\Bz-\Bzb$ mixing and decay and the cosine term is the
direct \CP\ asymmetry.

The \BB\ pair created in the \UfourS\ decay evolves coherently. Therefore, determining the flavor (\Bz\ or \Bzb)
of one $B$ meson ($B_{tag}$) also determines the flavor of the other $B$ at the time of the $B_{tag}$ decay. The second $B$ will continue to 
oscillate between flavor states.
The effective
tagging efficiency is $Q \equiv \sum_i {\eps_i (1-2\mistag_i)^2} = (30.4 \pm 0.3)\,\%$, where the 
sum over $i$ represents 6 mutually exclusive tagging categories, $\eps$ is the fraction of events
in each category, and \mistag\ is the fraction of incorrectly tagged events. The boosted center of mass 
allows for a measurement of the spatial separation of the $B$ meson decay vertices to be converted into the
proper time difference used in Eqn.~\ref{eq:acp}.

An unbinned, extended maximum likelihood fit (MLF) is used to separate signal events from background
and extract the \CP\ parameters, $S$ and $C$. Two kinematic variables from the \UfourS\ decay
are calculated and used in the MLF. They are the energy-substituted mass
$\mes=\sqrt{\frac{1}{4}s-\pvec_B^2}$
and energy difference $\DE = E_B-\half\sqrt{s}$, where
$(E_B,\pvec_B)$ is the $B$-meson 4-momentum vector, and
all values are expressed in the \UfourS\ rest frame. Event shape variables are used to distinguish
jet-like $q\bar{q}$ ($q=u,d,s,c$) events from nearly-isotropic $B$ meson decays. The
invariant mass and angular distribution (for vector mesons) are used to further separate
signal from background events. Background from \BB\ events tends to be small, and is included
as a component in the MLF where needed. Fits to Monte Carlo simulations
are used to determine signal and \BB\ PDF shapes. Fits to on-peak data
sidebands are used to determine the $q\bar{q}$ PDF shapes.

\section{\stwob\ from $\Bz\ra c\bar{c}K^0$}
The most precise measurement of $\beta$ comes from $\Bz\ra c\bar{c}K^0$ decays, where the
$b$ quark decays via the CKM-favored $V_{cb}$ transition to a $c \bar{c} s$ final state. In
these decays, $S_{\btoccbars} = -\eta_{f}\stwob$, where
$\eta_{f}$ is the \CP\ eigenvalue of the final state\cite{mishima}. A recent model-independent
calculation finds an expected deviation of $S_{\btoccbars}$ from $-\eta_{f}\stwob$ of $0.000\pm0.017$\cite{ciuchini}.

The measurement
presented here combines the results for several such final states: $\jpsi\KS(\pi^+\pi^-)$,
$\jpsi\KS(\pi^0\pi^0)$, $\psi(2S)\KS(\pi^+\pi^-)$, $\chi_{c1}\KS(\pi^+\pi^-)$, $\eta_c\KS(\pi^+\pi^-)$,
$\jpsi\KL$ and $\jpsi K^{*0}(\KS\pi^0)$. The result of the combined MLF is
\begin{equation}
\stwob\ = 0.710\pm0.034\pm0.019,
\end{equation}
where the first error is statistical and the second is systematic. This result, based on a 
data sample of 348 million \BB\ pairs, is consistent with the current world average
of $0.675\pm0.026$\cite{hfag}. Thus, $\beta$ is the most precisely measured CKM angle.

\section{\stwob\ from $b\ra s$ Penguins}
Decays of $\Bz$ mesons to charmless hadronic final states such as
$\etapr K^0$ proceed mostly via a single loop (penguin)
amplitude.  In the SM the penguin amplitude has approximately the same
weak phase as the $b \to c \bar{c} s$ transition, but it is sensitive to
the possible presence of new physics due to 
heavy particles in the loop\cite{Penguin}.
If the only contribution to these decays were from the dominant SM penguin processes,
$S_{b\ra s} = -\eta_{f}\stwob$ as in the \btoccbars\ case. 
However, other decay processes can contribute as well. Pollution from non-leading
order diagrams can cause $S_{b\ra s} \neq 
S_{\btoccbars}$, or $\Delta S \neq 0$, with $\Delta S \equiv
S_{b\ra s} - \stwob$. SM predictions and theoretical uncertainties 
for $\Delta S$ range from $\sim -0.05 - +0.20$ depending on the decay channel,
\cite{BN,BN2,Cheng,Zupan} where the cleanest modes have $\Delta S \sim 0.01\pm0.01$.
Any further deviation of $\Delta S$ from zero could be
due to the presence of new physics in the loop, which is not possible
in the \btoccbars\ case. Recent results from seven such $b\ra s$ 
penguin-dominated decay channels are shown in Table~\ref{tab:btos}.

\begin{table}[ph]
\tbl{Time-dependent \CP\ asymmetry parameter $S$ and data sample size
for $b\ra s$ penguin-dominated charmless $B$ decays. The first uncertainty
is statistical and the second is systematic.}
{\footnotesize
\begin{tabular}{lcc}
\hline\hline
Mode         & $ S$       &     \# \BB\ (millions)    \\
\hline
$B^0\ra\etapr K^0$~\cite{etapk} & \msp$0.58\pm0.10\pm0.03$  & 384 \\
$B^0\ra K^+K^-K^0$~\cite{kkks} & \msp$0.66\pm0.12\pm0.06$  & 347 \\
~~$B^0\ra \phi K^0$~\cite{kkks} &\msp$0.12\pm0.31\pm0.10$   & 347 \\
~~$B^0\ra f_0 K^0$~\cite{kkks} & \msp$0.35\pm0.34\pm0.08$   & 347 \\
$B^0\ra\omega\KS$~\cite{omks} & $0.62^{+0.25}_{-0.30}\pm0.02$  & 347 \\
$B^0\ra\rho^0\KS$~\cite{rhoks} & \msp$0.20\pm0.52\pm0.24$  & 227 \\
$B^0\ra\pi^0\KS$~\cite{piks} & \msp$0.33\pm0.26\pm0.04$ & 348 \\
$B^0\ra\pi^0\pi^0\KS$~\cite{pipiks} & $-0.72\pm0.71\pm0.08$  & 227 \\
$B^0\ra\KS\KS\KS$~\cite{ksksks} & \msp$0.66\pm0.26\pm0.08$  & 384 \\
\hline\hline
\end{tabular}\label{tab:btos}}
\vspace*{-10pt}
\end{table}

As shown in Table~\ref{tab:btos}, each measurement implies a negative value for
$\Delta S$. Moreover, the theoretical 
SM predictions for $\Delta S$ tend to be positive in nearly all cases.
However, the uncertainties, both experimentally and theoretically, are 
still sufficiently large that no definite conclusions can be reached at
this point. The single most precise measurement, that of $B^0\ra\etapr K^0$,
now shows a deviation of $5.5\sigma$ from zero, which is the first observation
of mixing-induced \CP\ violation in a charmless $B$ decay. The deviation from
\stwob, however, is $\sim1\sigma$. No individual channel represents a deviation
from \stwob\ $\gsim 2\sigma$.

\section{Measurements of \ctwob}
The measurement of \stwob\ leaves a 4-fold ambiguity in the value of $\beta$. This ambiguity 
can be partially resolved with a measurement of \ctwob. The final state
$\jpsi K^{*0}(\KS\pi^0)$ contains both \CP-even and \CP-odd components. A
full angular analysis of this final state allows for the extraction of 
\ctwob\ with the result $\ctwob > 0$ with 86\% confidence based on
a data sample of 88 million \BB\ pairs\cite{ctwob}.

Recently, two new technique have been used to deduce the sign of
\ctwob. Both $\Bz$ and $\Bzb$ mesons decay to the final 
state $\Dstarp\Dstarm\KS$. A potential interference effect of the decay proceeding
through an intermediate resonance can be measured by dividing the 
$B$-decay Dalitz plot into regions with
$m^2(\Dstarp\KS) >(<) m^2(\Dstarm\KS)$\cite{ref:browder}. The resulting measurement
concludes $\ctwob\ > 0$ with 94\% confidence\cite{BDDKS} in agreement with
the result from $\jpsi K^{*0}$. This result is based on a data sample
of 230 million \BB\ pairs.

A second new technique to determine the sign of \ctwob\ utilizes the decay
$B^0\ra D^0(\pi^+\pi^-\KS)h^0$, which can occur with or without $B^0-\bar{B}^0$
mixing, where $h^0$ represents an $\eta$, \etapr, $\pi^0$ or $\omega$ 
meson. Interference effects are visible across the $D^0$ Dalitz 
plot\cite{BDh-theory}. A full Dalitz plot fit measures $\stwob\ = 0.45\pm0.35\pm0.05\pm0.07$
and $\ctwob\ = 0.54\pm0.54\pm0.08\pm0.18$, where the first uncertainty is statistical, the second
is systematic and the third is theoretical, based
on a data sample of 311 million \BB\ pairs. This result shows a preference for
a solution of $\beta$ over $\pi/2-\beta$ with 87\% confidence\cite{BDh-exp}, in
good agreement with the previously reported measurements.

\section{Conclusions}
A variety of recent measurements of the CKM angle $\beta$ are reported
from \babar. The direct measurement of \stwob\ from \btoccbars\ channels
continues to be the most precise measurement. The sign of \ctwob\ is now
determined to be positive with at at least 86\% confidence in three independent 
measurements.  Several $b\ra s$ penguin-dominated charmless final
states continue to show a trend toward values of $S_{b\ra s} < \stwob$.


\begin{thebibliography}{0}
\bibitem{ref:ckm}
N.~Cabibbo, \jprl{10}, 531 (1963);
M.~Kobayashi and T.~Maskawa, Prog.\ Th.\ Phys.\ {\bf 49}, 652 (1973).

\bibitem{ref:babar}
\babar\ Collaboration, B.\ Aubert {\em et al.},
\nima{479}, 1 (2002).

\bibitem{ref:pep}
PEP-II Conceptual Design Report, SLAC-0418 (1993).

\bibitem{ref:babarprd}
\babar\ Collaboration, B.\ Aubert {\em et al.}, \jprd{66}, 032003 (2002).

\bibitem{mishima}
H-n.~Li, S.~Mishima, arXiv:hep-ph/0610120 (2006).

\bibitem{ciuchini}
M.~Ciuchini, M.~Pierini and L.~Silvestrini,
Phys.\ Rev.\ Lett.\  {\bf 95}, 221804 (2005).

\bibitem{hfag}
Heavy Flavor Averaging Group, E. Barberio \etal, arXiv:0704.3575v1[hep-ex] (2007).

\bibitem{Penguin}
Y.~Grossman and M.~P.~Worah, \plb{395}, 241 (1997);
D. Atwood and A. Soni, \plb{405}, 150 (1997);
M. Ciuchini \etal, \jprl{79}, 978 (1997). 

\bibitem{BN} M. Beneke and M. Neubert, \npb{675}, 333 (2003).
\bibitem{BN2} M. Beneke, \plb{620}, 143 (2005);
             G. Buchalla \etal, JHEP~{\bf 0509}, 074 (2005).  
\bibitem{Cheng}
H.~Y.~Cheng \etal, \jprd{72}, 014006 (2005), \jprd{71}, 014030 (2005); S. Fajfer \etal,
\jprd{72}, 114001 (2005).
\bibitem{Zupan}
A.~R.~Williamson and J.~Zupan, \jprd{74}, 014003 (2006).

\bibitem{etapk}
\babar\ Collaboration, B.\ Aubert {\em et al.}, \jprl{98}, 031801 (2007).

\bibitem{kkks}
\babar\ Collaboration, B.\ Aubert {\em et al.}, arXiv:hep-ex/0607112.

\bibitem{omks}
\babar\ Collaboration, B.\ Aubert {\em et al.}, arXiv:hep-ex/0607101.

\bibitem{rhoks}
\babar\ Collaboration, B.\ Aubert {\em et al.}, \jprl{98}, 051803 (2007).

\bibitem{piks}
\babar\ Collaboration, B.\ Aubert {\em et al.}, arXiv:hep-ex/0607096.

\bibitem{pipiks}
\babar\ Collaboration, B.\ Aubert {\em et al.}, arXiv:hep-ex/0702010.

\bibitem{ksksks}
\babar\ Collaboration, B.\ Aubert {\em et al.}, arXiv:hep-ex/0607108.

\bibitem{ctwob}
\babar\ Collaboration, B.\ Aubert {\em et al.}, \jprd{71}, 032005 (2005).

\bibitem{ref:browder}
T.E.~Browder {\em et al.}, \jprd{61}, 054009 (2000).

\bibitem{BDDKS}
\babar\ Collaboration, B.\ Aubert {\em et al.}, \jprd{74}, 091101 (2006).

\bibitem{BDh-theory}
A.~Bondar, T.~Gershon and P.~Krokovny, Phys.\ Lett.\  B {\bf 624}, 1 (2005).

\bibitem{BDh-exp}
\babar\ Collaboration, B.\ Aubert {\em et al.}, arXiv:hep-ex/0607105.

\end{thebibliography}
\end{document}